\documentstyle[twocolumn,aps,prl,epsf,floats]{revtex}
\begin{document} 

\title{Disorder Averaging and Finite Size Scaling}
\author{Karim Bernardet$^1$, Ferenc P\'azm\'andi$^{2,3}$ 
and G. G. Batrouni$^1$ }
\address{$^1$Institut Non-Lin\'eaire de Nice, Universit\'e de Nice-Sophia
Antipolis, 1361 route des Lucioles, 06560 Valbonne, France}
\address{$^2$Theoretical Physics Department, KLTE, H-4010
  Debrecen, P.O. Box 5, Hungary}
\address{$^3$Research Group of the Hungarian Academy of Sciences, 
Institute of Physics, TU Budapest, H-1521 Hungary} 

\address{\mbox{ }}

\address{\parbox{14cm}{\rm \mbox{ }\mbox{ } We propose a new picture
of the renormalization group (RG) approach in the presence of disorder,
which considers the RG trajectories of each random sample (realization)
separately instead of the usual renormalization of the
averaged free energy. The main consequence of the theory is that
the average over randomness has to be taken {\it after}
finding the critical point of each realization.
To demonstrate these concepts, 
we study the finite-size scaling properties of the
two-dimensional random-bond Ising model.
We find that most of the previously observed finite-size
corrections are due to the sample-to-sample fluctuation of the
critical temperature and
scaling is more adequate in terms of the new scaling variables.}}
\address{\mbox{ }} \address{\parbox{14cm}{\rm \mbox{ }\mbox{ } PACS
numbers: 75.10.Nr, 75.40.Mg, 05.70.Fh, 72.15.Rn }} \address{\mbox{ }}
\maketitle

\narrowtext

Finite size scaling (FSS) is a very powerful tool of theoretical
physics: it allows us to extract some properties of the infinite
system near a phase transition by studying finite, numerically
accessible samples. It is therefore of major interest to have a clear
theoretical background behind the bold extrapolation from finite to
infinite sizes. Though the basic concepts can be summarized in a few
lines, the theory of FSS is far from trivial \cite{cardy1} even for
clean systems.  Randomness brings in additional complexity, and a
deeper understanding of FSS in disordered systems is still
lacking. The main difference with the clean case is that somehow we
have to average over the different random samples.  There is an
on-going discussion of whether the way the disorder average is taken
influences the FSS results or not \cite{us,rieger,domany,harris2}, 
and if it does, what is
the ``correct'' average?  The importance of the details of averaging
is demonstrated most spectacularly by the so called Chayes et al
theorem \cite{chayes}, which claims that a certain finite-size
correlation length exponent cannot be smaller than $2/d$, $d$ being
the dimension of the disorder, for {\it any} phase
transition driven by quenched randomness.  It turns out \cite{us} that
the proof of this quite general statement relies entirely on the
specific manner the disorder was generated: a slight change in the
ensemble of the random samples gives a different final result. Further
studies along these lines showed that in some cases even the
numerically measured quantities do depend on the set of the disorder
realizations \cite{rieger,domany}, though there are claims that they
shouldn't \cite{harris2}.

In this Letter we propose to understand the role of the disorder based
on the scaling of a {\it single realization} instead of renormalizing
the averaged free energy. We argue that, eventhough the difference
between the two approaches is expected to vanish for infinite systems
and short-range interactions, it might be crucial for finite samples
and/or long-range forces. From a practical point of view, our main
result is that disorder averaging should be done {\it after} finding
the critical point of each sample independently.  We demonstrate how
this works in practice by performing an extensive numerical study of
the two-dimensional random-bond Ising model.

First, recall some basic ideas of FSS in clean systems. Close to a
continuous phase transition the correlation length, $\xi$, diverges as
$\xi(T)\sim\tau^{-\nu}$ with $\nu$ the correlation length critical
exponent and $\tau=|T-T_c|$ the distance from the critical 
temperature $T_c$ of the infinite system.
For a finite system, the size $L$ itself is measured in units of
$\xi$, i.e. a physical quantity $Q$ depends on $L$ only through
the ratio $L/\xi$, i.e.
\begin{equation}
       Q(T,L)=L^y \psi(L^{1/\nu}\tau),
\label{fss2}
\end{equation}
where $y$ describes the $L$-dependence at criticality.

The surprise in Eq.(\ref{fss2}) is that it contains the infinite 
system's correlation length (or critical temperature $T_c$), 
eventhough in a finite system the actual characteristic length
$\xi_L(T)$ is typically different from $\xi(T)$.  Indeed, while in
the high-temperature phase $\xi_L\sim\xi$, there is a temperature $T_c(L)$
where the correlation length reaches the system size, i.e. $\xi_L\sim L$. 
Below this temperature the whole sample becomes correlated
and $\xi_L$ is defined by subtracting this overall correlation.
We call the temperature $T_c(L)$, the critical temperature at
size $L$.
In terms of RG flows, the trajectories
bend towards high temperatures for $T>T_c(L)$, towards zero for
$T<T_c(L)$, and they ``stick around'' a fixed point for $T=T_c(L)$.

Admittedly, $T_c(L)$ is not a very well defined quantity, but the peak
in a susceptibility or specific heat may give it a sensible meaning.
Still, both the RG picture
and the behaviour of $\xi_L$ suggest that the scaling
variable of the problem is $\tau_L=T-T_c(L)$ instead of $\tau$,
leading to the FSS formula
\begin{equation}
       Q(T,L)=L^y f(L^{1/\nu}\tau_L). \label{fss1} 
\end{equation}
  
For clean systems the connection between Equations (\ref{fss2})
and (\ref{fss1}) is delivered by the scaling of $T_c(L)$:
\begin{equation}
T_c(L)=T_c+ C L^{-1/\nu}. \label{tc} 
\end{equation} The constant $C$ in this equation is {\it
not} universal, it depends e.g.  on the boundary conditions.
But once the details are fixed, C is constant for large $L$'s. 
Substituting Eq.(\ref{tc}) into Eq.(\ref{fss1}) gives us the usual
form of FSS (Eq.(\ref{fss2})).

Now we argue that in the presence of randomness, fixing the disorder
distribution and the boundary conditions is not enough to keep the
value of $C$ in Eq.(\ref{tc}) constant.  Due to the randomness, $C$
will fluctuate from sample to sample and under renormalization.
Consequently, $T_c(L)$ of a given disorder realization will fluctuate
as well, preventing the use of the infinite system's $T_c$ for all
samples and sizes, like in Eq.(\ref{fss2}).  At the same time $\tau_L$
remains a good scaling variable and, after an appropriate averaging,
Eq.(\ref{fss1}) holds.

The basic observation in support of the above is that the RG trajectories
for disordered systems are not smooth, but rather look like a random
walk. Each time we integrate out high-energy degrees of freedom, they
will contain some randomness. Accordingly, the
renormalized temperature will pick up a random part, too. Of course,
this fluctuation of the RG trajectory will scale as a negative
power of $L$, in the gaussian case as $L^{-d/2}$, and disappear
if $L\rightarrow\infty$. But in the case of FSS, we are comparing
temperatures as close as $\sim L^{-1/\nu}$, so 
for a $\nu$ close to $2/d$ the random walk
of the RG trajectories becomes important. Since $T_c(L)$ itself changes
under renormalization, we find that the critical surface 
will be random and different for each disorder realization.

\begin{figure}
\epsfxsize=3.0in
\epsfysize=2.25in
\epsffile{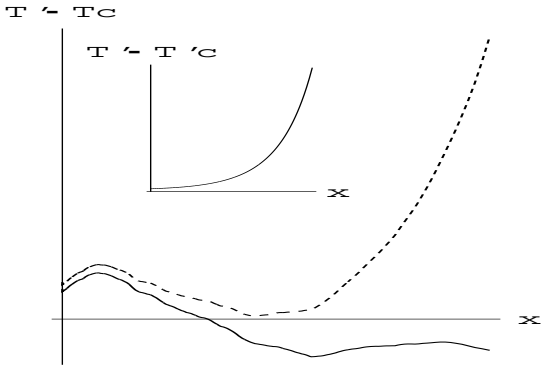}
\vskip 0.2cm
\caption{RG trajectories of temperatures for a single disorder realization
as a function of $x=\ln (L/L')$. The solid line represents $T'_c(L')$ and the
dashed line is for a close temperature $T'(L')$. The inset shows 
the difference of these two.}
\end{figure}

Now let's take a random sample of size $L$ and
consider the RG trajectory starting at a temperature $T$ close to 
the sample's $T_c(L)$.
After a renormalization step we get the renormalized values 
$L'$, $T'$, and $T'_c(L')$. According to the above arguments both
$T'$ and $T'_c(L')$ have a random part. But both $T'$ and $T'_c$
are temperatures, and they are close to each other, so it is
natural to suppose that their fluctuating part will be almost the same,
i.e. they are {\it correlated}. The main consequence of this
correlation is that $T'(L')-T'_c(L')$ will be a 
smooth function of $L'$, scaling with the exponent $1/\nu$, 
while $T'(L')-T_c$ will show the large fluctuations of the
random walk (see Fig.1). 

The standard (grand canonical \cite{us}) average uses $T_c$ only, and
completely neglects the correlations. Such an approach is justified as
long as the fluctuations of the RG trajectories are much smaller than
their distance from $T_c$. In the case of FSS, however, they might be
of the same order and the correlations become important: one has to
use $\tau_L=T-T_c(L)$ to extract the critical exponents. If $T$ is at
some distance (but not too far) from $T_c(L)$, so the sample contains
many correlated regions, the system is almost self averaging.  But
around $T_c(L)$ the remaining randomness in other quantities does not
necessarily scale to zero and, in order to use Eq.(\ref{fss1}), we
have to get rid of this extra noise by averaging.
 The ``correlated average'' \cite{us} requires then to find
the critical temperature of a given sample, and average over
realizations with the same $\tau_L$. In practice, this means
``shifting'' and superposing the curves of $Q(T)$ measured on
different random samples of the same size.  

\begin{figure}
\epsfxsize=3.0in
\epsfysize=2.25in
\epsffile{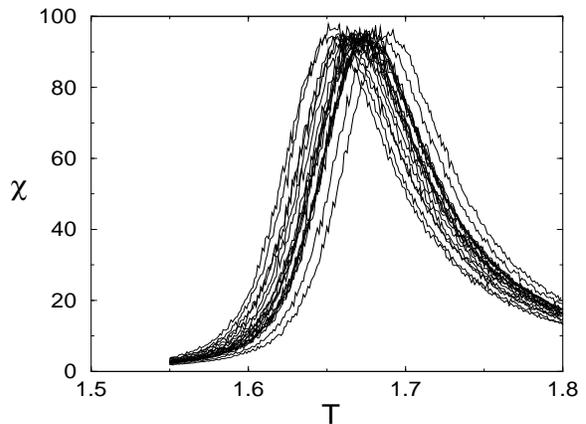}
\vskip 0.2cm
\caption{$20$ samples of susceptibility per spin versus temperature of
system of size $L=64$ with $r=0.5$.}
\end{figure}

We now test the above theoretical concepts on the two-dimensional
random-bond Ising model. We simulated $L\times L$ systems
($L=32,\dots.128$) with periodic boundary conditions using the Wolff
\cite{wolff} single-cluster algorithm to overcome critical slowing
down. Disorder was generated from a bimodal distribution: bonds had
two values, $J_1$ and $J_2$ (all positive) with equal
probabilities. The strength of randomness was tuned by changing the
ratio $r=J_1/J_2$ ($r=0.25,0.5$).  The exact critical temperature
$\beta_c=1/k_BT_c$ of this model is known \cite{fisch} as a function
of $r$ through : $\sinh(2\beta_cJ_2)\sinh(2\beta_crJ_2)=1$.  For each
measurement, we used up to $10^4$ Monte Carlo (MC) steps each
comprising $10$ cluster updatings and we used $10^4$ steps for
equilibration. To avoid inaccuracies \cite{ferrenberg} due to
unfortunate choice of the random number generator, we compared results
obtained from different generators. We concentrated on the
susceptibility, defined as
\begin{equation} 
\chi={1\over L^2}{\langle M^2 \rangle - \langle |M|\rangle^2 \over T},
\label{chi}
\end{equation} 
where $M$ is the total magnetization of the sample.

In Figure 2 we show the susceptibilities of different disorder
realizations of the same size ($L=64$) as a function of
temperature. We see large sample-to-sample fluctuations, exceeding
the thermal fluctuations by at least an order of magnitude. At the
same time, it is obvious that the curves are quite similar to each
other, they are just ``displaced''. This is exactly what we expect on
the basis of random RG trajectories: for each sample, $T_c(L)$ is
different, which explains the displacement of the curves.

\begin{figure} 

\epsfxsize=3.0in 
\epsfysize=2.25in 
\epsffile{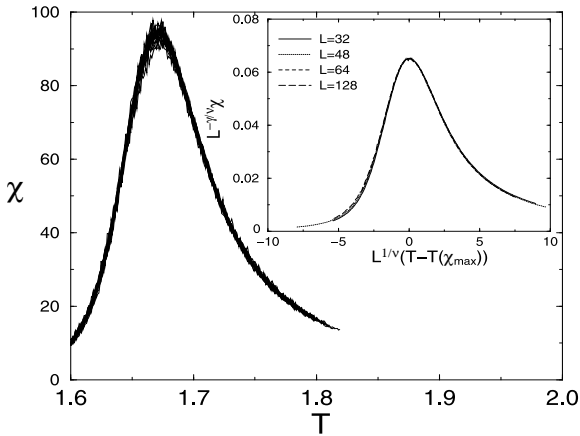} 
\vskip 0.2cm
\caption{The same $20$ realizations as in Figure 2 for $L=64$ and
$r=0.5$ after the shift. Inset: scaling with the correlated average
($\nu=1$, $\gamma=1.75$ for $L=32$ ($50$ real), $L=48$ ($20$ real), $L=64$ ($20$ real) and $L=128$ ($20$ real)). For each L, $T(\chi_{max})$ is the
temperature at the peak of the curve obtained after the correlated
average.}  
\end{figure}

To study the shape of the susceptibilities of the different disorder
realizations, in Figure 3 we ``shifted'' the curves to have each
sample at the same $T_c(L)$ (see below for details). We emphasize that
these are the very {\it same} data as in Fig. 2. The excellent overlap
of the different samples' susceptibilities demonstrates that
$\tau_L=T-T_c(L)$ is indeed the good scaling variable.  Fig. 3 also
shows that, as expected, disorder fluctuations are pronounced only at,
or around $T_c(L)$.  Our data indicate that the relative fluctuations
of the peak heights, are in the same order for all studied system
sizes, depending only on the disorder strength $r$.

In terms of
averaging over disorder, Figures 2 and 3 correspond to the grand
canonical and correlated averages, respectively: the latter achieves a
spectacular noise reduction, but it is still to see, which one
reproduces the expected scaling of the very large system.

Without randomness $\nu_{pure}=1$, and a perturbative RG approach
predicts that small disorder is marginally irrelevant \cite{cardy2}:
\begin{equation} 
{d\Delta \over dx}=-8 \Delta^2+{\cal O}(\Delta^3), \label{delta}
\end{equation} 
where $\Delta$ is proportional to the square dispersion of the random
bonds, and $x\propto \ln(L^{-1})$.  According to Eq.(\ref{delta}), the
disorder scales to zero, but only logarithmically with $L$, so we have
to take it into account in the RG equations of other quantities, like
the reduced temperature $\tau$, \begin{equation} {d\tau \over dx}=(1-4
\Delta)\tau+\dots.  \label{tau} \end{equation} This equation predicts
an effective exponent $\nu_{eff}\sim 1+4\Delta$, which approaches
$\nu_{pure}=1$ very slowly. Since the randomness {\it does not couple}
to the magnetization in first order, one expects that the
susceptibility exponent $\gamma/\nu=1.75$ remains {\it unchanged}.
Even though these results were obtained by using replicas and
grand-canonical disorder average, for a short-range-interaction model
and very large system sizes we still expect them to be correct.

The detailed form of the above scaling corrections is still under
debate even today \cite{brazil,kim}. Here we wish to concentrate only
on their qualitative nature: disorder introduces corrections to $\nu$
(the width of the susceptibility peak), but {\it not} to $\gamma/\nu$
(the height of the peak at criticality). As we will see, this
expectation is satisfied {\it only} with the correlated average.

The major difficulty of the correlated average is to find $T_c(L)$ of
a given disorder realization, and ``shift'' the different samples'
curves as in Fig. 3.  Trying to identify the peak of the
susceptibility for each realization is one possibility \cite{domany},
but both thermal and random fluctuations are biggest at this
point. Instead, we used the entire susceptibility curves and minimized
the ``distance''\cite{norm} between them.  We verified that the final
results do not depend on the details of this procedure, and the
average critical point $\overline{T_c(L)}$ scales to the exact $T_c$
when $L\rightarrow\infty$. The exact values are $T_c=1.641018$ for
$r=0.5$ and $T_c=1.239078$ for $r=0.25$. Our extrapolated
$L\rightarrow\infty$ numerical results are $T_c=1.640(1)$ and
$T_c=1.239(1)$ respectively.

In the case of small disorder, $r=0.5$, we found corrections to
scaling in the case of grand canonical average both for $\nu$ and
$\gamma/\nu$, though both of these corrections are relatively
small. This violates what is expected for $\gamma/\nu$. On the other
hand, for the available sizes, the correlated average gives an almost
perfect scaling plot with the {\it pure exponents}, as shown in Figure
3 (inset). No corrections were visible here.

\begin{figure}
\epsfxsize=3.0in
\epsfysize=2.25in
\epsffile{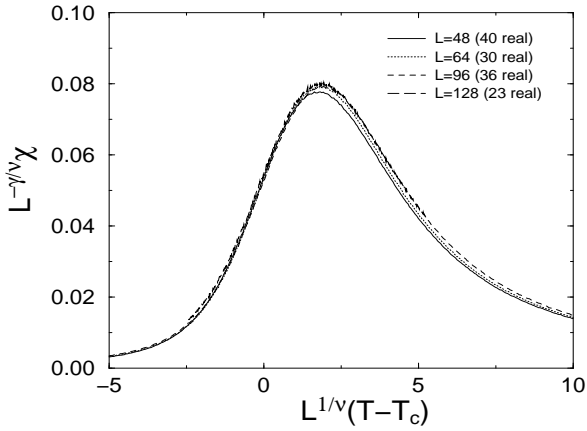}
\vskip 0.2cm
\caption{Grand canonical average, $r=0.25$ ($\nu=1$, $\gamma=1.75$).}
\end{figure}

The differences between the two disorder averages are even more
pronounced for stronger disorder, $r=0.25$. Clearly, for the grand
canonical average (Fig. 4) not only the widths but also the heights of
the peaks show sizable corrections to scaling. Note that there are
{\it no} corrections for the heights (scaled by $\gamma/\nu$) when the
data have been evaluated with correlated average (Fig. 5). For this
disorder strength the corrections in $\nu$ already appear, and an
effective thermal exponent $1/\nu_{eff}\sim 0.92$ gives a good
description of the data within this range of sizes (see the inset of Fig. 5). We
emphasize that only the results of the correlated average reproduce
our expectations for the infinite system scaling.

\begin{figure}
\epsfxsize=3.0in
\epsfysize=2.25in
\epsffile{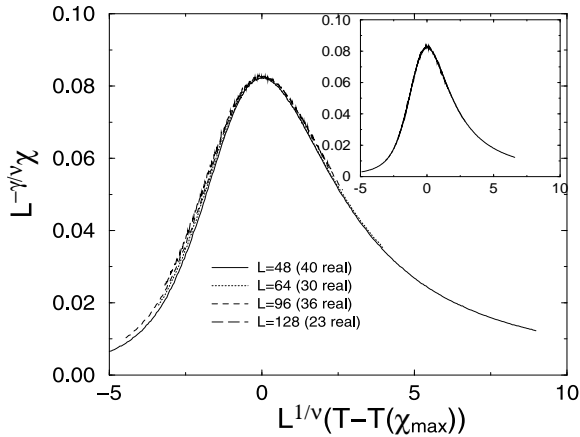}
\vskip 0.2cm
\caption{Correlated average, $r=0.25$ ($\nu=1$ and
$\gamma=1.75$, inset: $1/\nu\sim0.92$ and $\gamma/\nu=1.75$).}
\end{figure}

In addition to the susceptibility, other singular quantities, like the
specific heat, show critical behaviour. A question of consistency
arises: Do the {\it same} temperature shifts calculated from the
susceptibilities of different samples give the best collapse of the
specific heat curves?  Indeed, the answer is yes, as can be seen in
Fig. 6. This supports our theory that $\tau_L=T-T_c(L)$ of a given
sample is a good scaling variable for any critical quantity.

\begin{figure}
\epsfxsize=3.0in
\epsfysize=2.25in
\epsffile{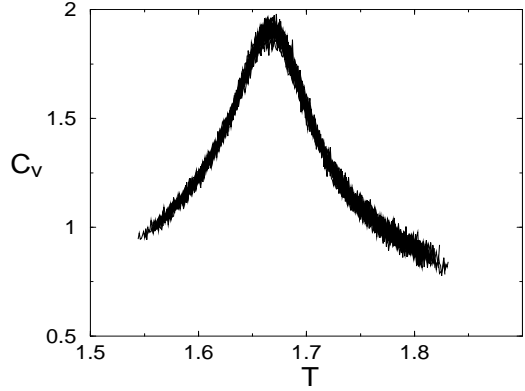}
\vskip 0.2cm
\caption{Collapse of the specific-heat curves using the same 
shifts of temperature as in Figure 3 ($r=0.5$, $L=64$).}
\end{figure}

We have proposed a new picture of the RG in random systems, which
leads to a recently introduced way of disorder averaging for FSS, the so
called ``correlated average''\cite{us}.  We studied in detail the
FSS properties of the $d=2$ disordered Ising model, and
found that only the correlated average reproduces the expected
behaviour of the susceptibility, in addition to spectacular noise
reduction in averaged quantities. A detailed account of our
simulations' results will be published elsewhere.

We would like to thank Alex Hansen (Trondheim) and the CNUSC in
Montpellier for computer time. FP thanks INLN (Nice), the Hungarian
Science Foundation (OTKA29236), and the B\'olyai Fellowship for
financial support.


\begin{references}
\bibitem{norm} To measure the distance of two functions $f_1$ and $f_2$,
we used the norm $\int dx |f_1(x)-f_2(x)|$.
\end{references}
\end{document}